**Lemaître's Limit**

*by Ian Steer, Toronto Centre*
*NASA/IPAC Extragalactic Database of galaxy Distances,*
*Pasadena, California, USA*
*(iansteer1@gmail.com)*

Georges Lemaître gave a theoretical proof, for his 1927 doctoral thesis in astronomy, that the "maximum spherical radius" of our Universe can be computed from first principles to be 14.2 billion light-years (Lemaître 1927a). That estimate, which is known as Lemaître's limit, is based on Lemaître's dynamic-equilibrium theory of the Universe. It is surprisingly close to current estimates of the Universe's age. That age has been firmly established at approximately 14 billion years, based on multiple measurements, including measurements of the extragalactic distance scale by the NASA *Hubble Space Telescope* Key Project (Freedman *et al.* 2001), and of the cosmic microwave background radiation by the NASA *Wilkinson Microwave Anisotropy Probe* in combination with measurements of the distribution of galaxies by the Sloan Digital Sky Survey (Tegmark *et al.* 2004). Recently released final results from the full nine years of measurements by the *Wilkinson Microwave Anisotropy Probe* put the Universe's age at 13.74 ± 0.11 billion years (Bennett *et al.* 2013).

It is surprising that Lemaître's limit has been all but forgotten. Such coincidence, to within 3 percent, between the predicted size and observed age of the Universe ought to be of interest. Yet Lemaître's limit, his dynamic-equilibrium theory that predicted that limit, and other results from his earliest cosmological research are all but unknown to modern science. Only a single reference could be found, on a search of the NASA Astrophysics Data System, to Lemaître's thesis (Lemaître 1927a). By contrast, Lemaître's expanding-Universe theory is well recognized (Lemaître 1927b) and it is for that theory that he is considered a founding father of Big Bang cosmology and why today's standard cosmological model is known as the Friedmann-Lemaître-Robertson-Walker universe. Lemaître was the first to provide solid theoretical evidence for the expansion of the Universe. He even calculated the Hubble constant of expansion, two years before Hubble, which occurred only after Hubble had uncovered observational proof of expansion (Hubble 1929) [see "Who discovered expansion", following].

Lemaître's limit might come back into modern astronomy, as did the cosmological constant. Indeed, the coincidence it represents between the size and age of the Universe has become more meaningful since the resurrection in the late 1990s of the cosmological constant (also known as vacuum or "dark" energy). That resurrection was based on observations of distant Type Ia supernovae, the same work that earned Riess, Schmidt, and Perlmutter the 2011 Nobel Prize in physics (Riess *et al.* 1998, Perlmutter *et al.* 1999). Verification of the cosmological constant has restored the relationship between the Universe's age in years and its size in light-years. Without the cosmological constant, expanding theories such as Friedmann's estimate the expansion age of the Universe as only 2/3 of the light-travel time required to reach the Hubble expansion radius (Friedmann 1922). The Universe's age could not coincide with Lemaître's limit to better than 33 percent. With the cosmological constant, the Universe's expansion age of 2/3 of the light-travel time is divided by 0.7, the estimated fraction of the Universe's total energy density attributable to the cosmological constant. As a result, age and distance in today's standard model once again equal one another to within 5 percent, *i.e.* to within 0.666/0.7 = 0.95. In essence, the cosmological constant restores the relationship that originally existed, where ages in years and distances in light-years were equivalent and interchangeable. In the earliest expanding theories, including de Sitter's and in Lemaître's dynamic-equilibrium theory, there was a one-to-one relationship between the expansion age of the Universe and the distance light has travelled since expansion began (de Sitter 1917, Lemaître 1927a). Lemaître's limit and the Universe's age coincidence, therefore, is of more interest now than it might have been historically because of the restoration of the cosmological constant.

Differences between Lemaître's dynamic-equilibrium and expanding theories of the Universe are shown in Figure 2. Note the dynamic-equilibrium theory is a hybrid. It incorporates into one theory effectively all of the probable theories possible according to Einstein's general theory of relativity. Those include both dynamic and non-dynamic theories, including expanding and/or contracting theories, as well as static theories. As a result, Lemaître's dynamic-equilibrium Universe includes more than simply the expansion radius of the expanding theories, as the figure shows. It also includes the Einstein radius of the static theory as an inner boundary, and the Schwarzschild radius as an outer boundary. The Schwarzschild radius, which is the radius of a black hole's event horizon, is usually taken to define the horizons of objects within the Universe rather than the horizon of and exterior limit to the Universe itself. In comparison, Lemaître's expanding theory can be summarized by the expansion radius alone, as shown separately. That radius, described by Lemaître as the de Sitter radius, is now defined as the Hubble radius.

Lemaître's dynamic-equilibrium theory, as a hybrid, incorporates multiple theories, their multiple radii, and their multiple possibilities. Basically, he is offering a sphere-within-sphere theory, similar to the earlier Wright Universe (Wright & Rafinesque 1837). Further, rather than simply an expanding Universe with Hubble's radius and/or a static one with Einstein's, Lemaître's dynamic-equilibrium theory and to a first approximation, Lemaître's limit, offers a Universe with boundaries that limit the expansion radius. In expanding theories, that radius can reach any size up to and including infinitely large values. In the dynamic-equilibrium theory, however, the expansion radius is limited to expanding (and/or contracting) between inner and outer boundaries, as shown. Those boundaries are defined as noted, inwardly by the Einstein static radius and outwardly by the Schwarzschild event-horizon radius. In other words, the Universe might exist within a black hole. That is no longer a unique or original view. Its origin, however, can be traced to Lemaître's dynamic-equilibrium theory. That theory, though all-encompassing, was nevertheless abandoned by Lemaître after Hubble discovered observational proof of expansion (Hubble 1929). Thereafter, Lemaître pursued his purely expanding theory. In the process, the maximum spherical radius was replaced by one to be determined by observation and all but forgotten.

Lemaître's dynamic-equilibrium theory might well be relevant to today's cosmologists, precisely because of its all-encompassing hybrid nature. It incorporates purely dynamic and expanding theories by placing them in dynamic equilibrium. By assuming balance or equilibrium between gravitational attraction and electric repulsion *ad hoc*, just as Einstein did with his first formulation of general relativity but in a static theory, Lemaître is including the cosmological constant in expanding theories. Inclusion of the cosmological constant in today's expanding theory is the reason the Friedmann-Robertson-Walker standard model (before confirmation of the cosmological constant) became the Friedmann-Lemaître-Robertson-Walker model. In contrast to his purely expanding theory (Lemaître 1927b), however, Lemaître's dynamic-equilibrium theory (Lemaître 1927a) also incorporates non-dynamic and purely static theories by framing those as stationary theories. That allows static theories to feature important properties of dynamic theories including expansion and/or contraction, while also still retaining the all-important properties of the cosmological constant. Einstein later disavowed the cosmological constant. He called it his biggest blunder precisely because it was *ad hoc*, and after learning of Friedmann's expanding theory of 1922 and then of Hubble's observational confirmation of expansion in 1929, was relieved to drop it (Friedmann 1922, Hubble 1929, Einstein 1945). Today, however, the cosmological constant is the only physical mechanism that is both understood and observationally confirmed to be able to counter-balance gravity, as originally intended by Einstein and later Lemaître.

All three radii in Lemaître's dynamic-equilibrium theory were established within the first two years of Einstein's earliest formulation of the general theory of relativity in 1915 (Einstein 1915). They were found by Schwarzschild (1916), Einstein (1917, translated 1922, Eq. 124), and de Sitter (1917). Of these, Schwarzschild's radius of the event horizon of black holes is the most recognized. That radius is derived by an equation that is probably the most cited in astronomy after $E = mc^2$, namely $r_S = 2Gm/c^2$. The Schwarzschild radius of the Sun is widely understood to be 3.0 kilometres, based on the Newtonian gravitational constant (6.67380 x $10^{-8}$ cm$^3$/gm sec$^2$), the speed of light (2.99792458 x $10^{10}$ cm/sec)[1] and the Sun's mass (1.988435 x $10^{33}$ gm, Gundlach & Merkowitz 2000). Less widely recognized are the other radii, and in particular the fact that both are so closely related to Schwarzschild's radius, as the figure shows. Einstein's static-theory radius is derived by exactly the same equation as Schwarzschild's, excepting only for being smaller by a factor of $\pi$. So too, the Hubble expansion radius is derived by exactly Schwarzschild's equation, excepting only for it being smaller by a factor of 2.

Lemaître was the first to establish that all three radii ($r_S$, $r_E$, and $r_H$), then thought separate and unrelated solutions, might actually be three closely related sizes surrounding the same constant mass. The virial mass of expanding theories is the same as the virial mass of static theories, as first noted by de Sitter (1917), then Hubble (1926), and later Eddington (1930) and Einstein (1945). The virial mass is defined as the exact mass a larger body such as the Sun must have to prevent the gravitational escape of a smaller body such as the Earth, *i.e.* to overcome the smaller body's independent velocity of motion. For ultra-massive bodies such as the Universe, with smaller bodies such as galaxies having escape velocities approaching the speed of light, the exact virial mass required to prevent escape is also known as the gravitational mass.

Lemaître's limit is given explicitly in a formula, which is reproduced here in Equation 1 [2]. In that formula, Lemaître multiplies three terms together, the Einstein gravitational constant ($\kappa$), the square of the maximum virial radius ($r^2$), and the "invariant mass density" (d), which he writes as $8\pi\, a^2\, d$. Lemaître's limit, defined by that formula, occurs when the product of those terms reaches unity or one, or in other words, when those terms are mathematically in balance, with the Newtonian gravitational constant, G, and the velocity of light, c.

$$\kappa r^2 \rho = \left(\frac{8\pi\, G}{c^2}\right) r^2 \left(\frac{m}{8\pi\, r^3}\right) = \frac{G}{c^2}\frac{m}{r} = 1 \tag{1}$$

The Lemaître limit formula reduces to only two terms; the $r^2$ term cancels since density equals $m/r^3$, as shown in (1). Reduced to two terms, the formerly unknown mass/radius ratio of the Universe (m/r), taken as the first term, then becomes known, because it equals a known constant ratio ($c^2/G$), taken as the second term, as shown in (2).

$$\frac{m}{r} = \frac{c^2}{G} = 1.35 \times 10^{28} \text{ gm/cm} \tag{2}$$

Based on that constant, universal, and unifying ratio ($c^2/G$), Lemaître makes only two assumptions. First, our system of units for length, mass, and time holds true for the Universe (note Lemaître uses centimetres, grams, and seconds), and second, that the virial mass-radius relation also holds true. That is, the relation between the virial mass as defined earlier, and the corresponding escape radius from that mass for smaller bodies moving at the limiting velocity of light ($m = rc^2/G$) has universal application. Then, relative to a minimum radius that he defined as 1 cm, Lemaître computed a natural limit for the maximum radius. That limit is reached when the maximum-to-minimum-radius ratio itself reaches the same unifying ratio ($r_{max}/r_{1cm} = c^2/G$), as shown in (3a). Lemaître's limit is reached at a maximum radius of $1.35 \times 10^{28}$ cm (14.2 billion light-years), when the square of the Universe's virial radius equals its own virial mass in conventional units ($r^2 = m$), and when the virial radius equals the virial mass in natural units ($r = m$, where $c = G = 1$). Only at that limiting boundary radius are the mass, mass/radius ratio, and virial mass-radius relation all linked by the same unifying ratio, as shown in Eq. 3b.

$$r_{max} = \frac{r_{max}}{r_{1cm}}(1.0 \text{ cm}) = \frac{c^2}{G}\left(1.0\frac{cm^2}{gm}\right) = \frac{m}{r}\left(1.0\frac{cm^2}{gm}\right) \tag{3a}$$

$$r_{max} = 1.35 \times 10^{28} \text{cm} = 14.2 \text{ Gly}$$

$$m = \frac{m^2}{r^2}\left(1.0\frac{cm^2}{gm}\right) = r\frac{c^2}{G} = \frac{c^4}{G^2}\left(1.0\frac{cm^2}{gm}\right) \tag{3b}$$

$$m = 1.81 \times 10^{56} \text{gm} = 9.12 \times 10^{22} M_{Sun}$$

The unifying ratio, namely $c^2/G$, is simply the reciprocal of Einstein's constant, κ, without the 8π geometrical factor, since κ = $8πG/c^2$, as shown in (1).

Lemaître's limit lives on, though it is now known rather obliquely as the "Newtonian" or classical limit of general relativity. As recently as 2001, a version of Lemaître's limit was used to estimate the maximum radius of the Universe to four-digit precision, finding 13.83 billion light-years (Nowakowski 2001). That estimate, however, assumed vanilla values of the Hubble constant and the total density-to-critical density ratio (H = 100 km/s/Mpc, and $d_T/d_C$ = 1). If Lemaître's formula values are employed instead, for a Hubble constant and velocity-distance relation at the limits of the velocity of light, c, and the maximum radius, $r_{max}$ (where $H_L = c/r_{max}$ = 68.7 km/s/Mpc), with a total density of double the critical density (where $d_T/d_C$ = 2), the result then is Lemaître's maximum radius of 14.2 billion light-years, i.e. 14.2 = 13.8 × [($H_{100}/H_L$)/√2].

Lemaître's formula, if not the exact limit, evidently has currency in modern physics. Three facts, however, have been all but lost regarding what is nowadays referred to as the Newtonian limit. First, describing that limit as Newtonian or classical is incomplete at best, because it could scarcely have been foreseen, let alone foretold, before the advent of general relativity. Second, Lemaître is the first physicist known to have established its theoretical existence, yet his 1927 discovery and prediction remain unheralded. Third, the exact limit itself has been completely lost to modern physics. That is surprising. Lemaître's limit and observational estimates of the Universe's age, as said, coincide to within 3 percent.

Lemaître's name appears in the abstracts of more than one thousand astronomical papers as of the beginning of 2013, according to the joint NASA Smithsonian Astrophysical Observatory Astrophysics Data System. Yet, of the 45 papers found naming Lemaître in their abstracts in *The Astrophysical Journal*, none cites Lemaître's 1927 thesis. Of the subset of all papers searched naming Lemaître in their abstracts, some 120 papers or 10 percent of the total available, only 1 was found citing Lemaître's thesis, a review by Eisenstaedt (1993). That review divulges the existence of, but not the physics of, Lemaître's limit. Only one other reference to Lemaître's limit could be found in the modern literature, and that only after a pre-print of this JRASC manuscript was circulated on the astrophysics paper e-print archive online (http://arxiv.org/abs/1212.6566). A private communication, from Lemaître biographer Dominique Lambert, reveals that his biography of Lemaître includes seven pages referencing Lemaître's limit (Lambert 2000). Although published *en français*, an English translation is eagerly anticipated this year. Further interest in Lemaître's limit, beyond pedagogical, will depend on future observation-based findings regarding the Universe's size and age.

For astronomers to "discover" whether Lemaître's limit is true or mere coincidence will require, by definition, observational measurements with an accuracy of at least three sigma or 0.3 percent! Estimates of the Hubble constant, from which the Universe's age and size are derived, are accurate at present to within 3 percent. Those include the most recent results from the Carnegie Hubble Program, co-led by former NASA Key Project co-leaders Freedman & Madore (Freedman *et al.* 2012), and the Supernovae H0 Equation of State team, co-led by Reiss, 2011 Nobel Prize co-winner, and Macri (Riess *et al.* 2011). These programs, however, and others ongoing and planned, are aimed at achieving 1-percent accuracy. Following launch of the NASA *James Webb Space Telescope* in 2018, Hubble-constant estimates of that degree of accuracy might well be achieved. Aiding in that endeavour, within the next decade, giant ground-based telescopes with apertures of more than 30 metres will become available, including the Giant Magellan Telescope, the Thirty-Meter Telescope, and the European Extremely Large Telescope.

More down-to-Earth, poor-man's avenues to high-accuracy cosmological research now exist. Statistical and theoretical analyses can be conducted by anyone, thanks to the vast volumes of data that are already available and openly accessible to all. Examples include analyses of the database of Hubble-constant estimates published from 1927 to 2010 totaling 600 values, compiled by Huchra for the NASA *Hubble Space Telescope* Key Project and available online (Huchra 2010). Earlier analyses of subsets of those estimates led researchers to find a mean of H = 67 km/s/Mpc with a standard deviation of 5 percent (Gott, Vogeley, Podariu, & Ratra 2001), and then with additional estimates H = 68 km/s/Mpc with a standard deviation of only 2 percent (Chen, Gott, & Ratra 2003). An unpublished mean found by this writer based on 365 of the 487 estimates available in 2005, excluding 12 estimates published prior to 1960 and 110 based on non-local distance indicators, including gravitational lens and Sunyaev-Zeldovich effect-based indicators, resulted in H = 68.9 km/s/Mpc. That, coincidentally, is within 0.3 percent of the predicted Hubble constant given earlier based on Lemaître's limit, $H_L$ = 68.7 km/s/Mpc. The most precise Hubble constant to date, H = 69.32 km/s/Mpc based on the full nine-years of Wilkinson Microwave Anisotropy Probe measurements, is within 1% of the value based on Lemaître's Limit (Bennett *et al.* 2013).

Analysis of the NASA/IPAC Extragalactic Database of Galaxy Distances is another way to obtain theoretical yet highly accurate cosmological research. That database features essentially all of the redshift-independent extragalactic distance estimates published since 1980 and upon which most current estimates of the Hubble constant of proportionality between distance and velocity are based. It is co-led by this writer in collaboration with Madore, RASC member and annual contributor to the RASC *Observer's Handbook*, co-founder of the NASA/IPAC Extragalactic Database, former co-leader of the NASA *Hubble Space Telescope* Key Project, and current co-leader of the Carnegie Hubble Program.

In theory, with 60,000 redshift-independent distance estimates available for 12,000 galaxies, the Hubble constant could be found with an accuracy of better than 1 percent, based on having more than 2 orders of magnitude more distance measurements than the Key Project did in 2001, which achieved 10-percent accuracy. That estimate was based on 200 distance measurements for 100 galaxies, as compiled for the Key Project in the first *Hubble*-era database of extragalactic distance estimates (Ferrarese 2000).

One immediate example of armchair results based on "big data" analysis is a statistically derived estimate of the distance to the Large Magellanic Cloud galaxy made by this writer. That distance represents the anchor or zero point of the extragalactic distance scale. Based on analysis of 530 measurements available in the distances database, the accuracy of the calculated estimate is claimed to be 1.2 percent.

For now, reaction to Lemaître's limit depends mostly on one's views. Without question, it involves an unproven coincidence that was abandoned by its originator and has no place in today's standard model. Then again, the same acceptance of expansion in 1929 that caused Lemaître to abandon his dynamic equilibrium theory also caused Einstein to abandon his cosmological constant, since reborn. Might Lemaître's early ideas also be revived in a future standard model? That question is timely. Another recent estimate of the age of the Universe, based on the abundance of heavy-chemical elements observed in an extremely metal-poor K-type giant star, is 14.2 billion years. That was reported by Christopher Sneden, former editor of *The Astrophysical Journal Letters* (Sneden *et al.* 2003), and quoted in the popular press (see *Astronomy* magazine June 2005, p. 46, by Steve Nadis). If Lemaître's Limit is resurrected, it will revolutionize cosmology.

**Endnotes**

1 U.S. National Institute of Standards and Technology, Committee on Data for Science and Technology recommended values of the fundamental physical constants (Mohr, Taylor, & Newell 2012).
2 That formula originally appeared on page 23 of Lemaître's thesis, at the beginning of Section V, "Interpretation of the results" and shown in his Table V, column 2. The thesis is available on-line through the Massachusetts Institute of Technology (http://mit.dspace.org/bitstream/handle/1721.1/10753/36897534.pdf?sequence=1)

*Ian Steer followed an independently pursued avocation as a theoretical cosmology researcher from a non-astronomy background. In 2005, he was contracted by the California Institute of Technology as co-leader of the NASA/IPAC Extragalactic Database of Galaxy Distances. He now collaboratively pursues his vocation – to research, gather, and make publicly available data that astronomers use in extragalactic studies and to estimate cosmological parameters.*

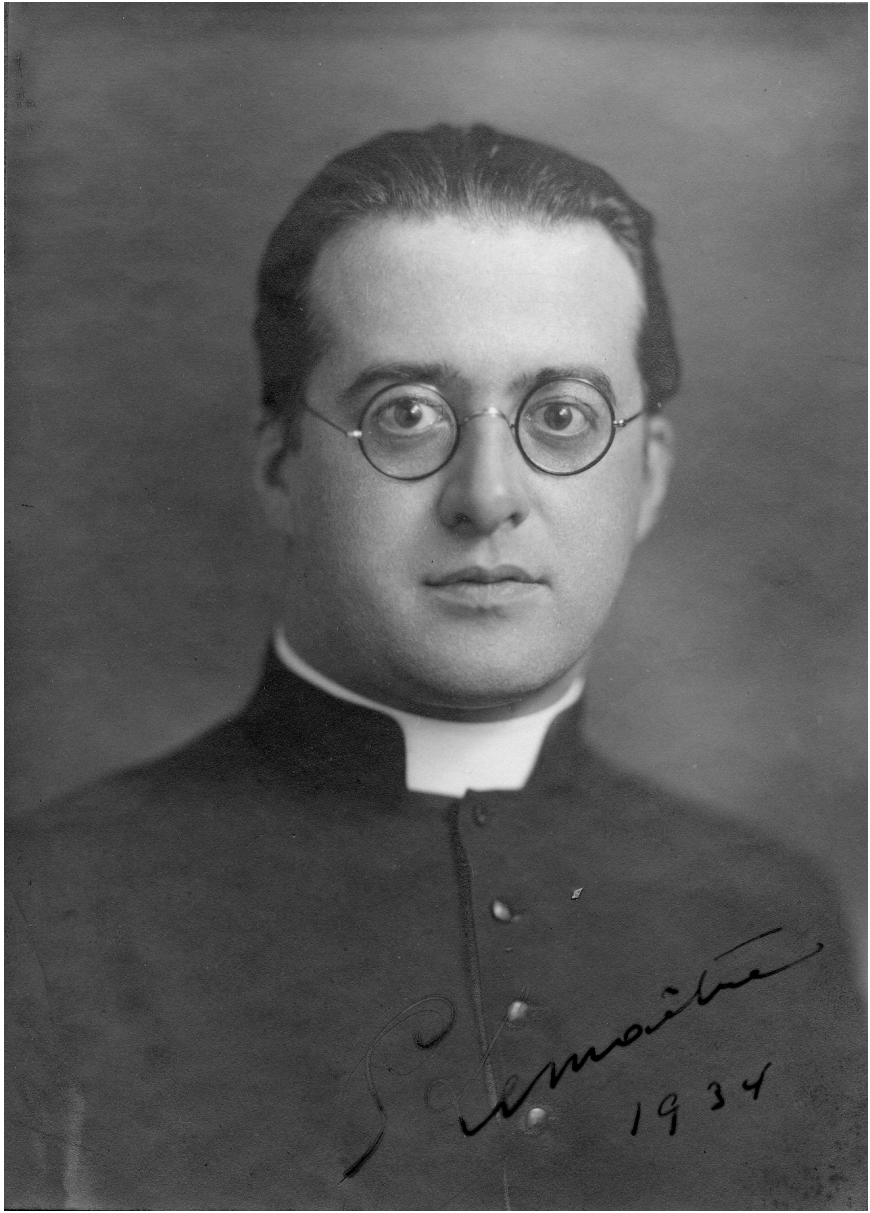

Figure 1 — Mgr. George Lemaître. Image courtesy Archives Georges Lemaître

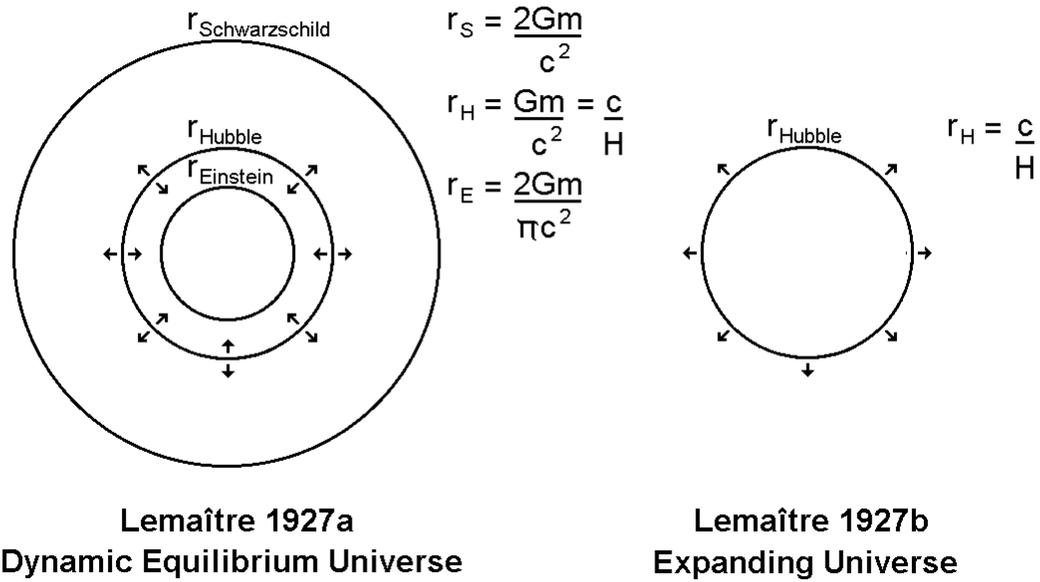

**Lemaître 1927a**
**Dynamic Equilibrium Universe**

**Lemaître 1927b**
**Expanding Universe**

Figure 2 — Two theories of the Universe: one nearly lost to modern physics (1927a), and one that forms the basis of today's standard cosmological model (1927b).

Who discovered expansion?

Much debate has ensued recently over who deserves credit for being first to discover the Universe is expanding. Lemaître's theoretical discovery of expansion in 1927 was not translated into English until two years after Hubble's observational discovery of expansion in 1929 (Lemaître 1927b, translated 1931). Further, that translation omitted the Hubble constant of expansion that Lemaître calculated in 1927. That omission, only recently revealed in these pages by frequent contributor to *JRASC* Sidney van den Bergh of the Dominion Astrophysical Observatory, peaked the debate (see van den Bergh 2011, *JRASC* **105,** 151). At least a dozen papers regarding Hubble's priority in the discovery of expansion have been published in the last two years alone, more papers regarding Hubble's priority than were written in all the years since 1929 combined. From *JRASC* pages and those papers, debate has leapt to the level of *Nature*. For example, van den Bergh continues to support Lemaître as the discoverer of expansion (see van den Bergh 2011, *JRASC* **105,** 197-199, and van den Bergh 2011, *Nature* **480,** 321). However, Mario Livio of the Space Telescope Science Institute has shown that Hubble's observational confirmation of expansion should not be tarnished. Livio found a letter by Lemaître proving that it was Lemaître himself who omitted the 1927 Hubble constant calculation from his 1931 translation (see Livio 2011, *Nature* **479,** 171-173). Your present writer has also supported Hubble's priority, including in these pages regarding Hubble and Swedish astronomer Knut Lundmark, and by noting that Lemaître discovered theoretical evidence for expansion, while Hubble discovered observational proof (see Steer 2011, *JRASC* **105,** 18-20, and Steer 2012, *Nature* **490,** 176).